\documentclass[10pt,preprint]{aastex} 

\shorttitle{Mg isotopes in Halo Dwarfs}
\shortauthors{Mel\'endez \& Cohen}

\begin{document}
\title {Magnesium Isotopes in Metal-Poor Dwarfs, the Rise of AGB Stars
and the Formation Timescale of the Galactic Halo\altaffilmark{1}}

\newcommand{\teff}{T$_{\rm eff}$ }
\newcommand{\tsin}{T$_{\rm eff}$}
\newbox\grsign \setbox\grsign=\hbox{$>$} \newdimen\grdimen \grdimen=\ht\grsign
\newbox\simlessbox \newbox\simgreatbox
\setbox\simgreatbox=\hbox{\raise.5ex\hbox{$>$}\llap
     {\lower.5ex\hbox{$\sim$}}}\ht1=\grdimen\dp1=0pt
\setbox\simlessbox=\hbox{\raise.5ex\hbox{$<$}\llap
     {\lower.5ex\hbox{$\sim$}}}\ht2=\grdimen\dp2=0pt\def\simgreat{\mathrel{\copy\simgreatbox}}
\def\simless{\mathrel{\copy\simlessbox}}
\newbox\simppropto
\setbox\simppropto=\hbox{\raise.5ex\hbox{$\sim$}\llap
     {\lower.5ex\hbox{$\propto$}}}\ht2=\grdimen\dp2=0pt
\def\simpropto{\mathrel{\copy\simppropto}}

\author{Jorge Mel\'endez} 
\affil{Research School of Astronomy \& Astrophysics, 
Australian National University, Mt. Stromlo Observatory, Weston ACT 2611,
Australia} \email{jorge@mso.anu.edu.au} 

\and

\author{Judith G. Cohen}
\affil{Palomar Observatory, MC 105--24, California Institute of Technology, Pasadena, CA
91125}\email{jlc@astro.caltech.edu}

\altaffiltext{1}{The data presented herein were obtained at the W.M. Keck Observatory, 
which is operated as a scientific partnership among the California Institute of Technology, 
the University of California and the National Aeronautics and Space Administration.}

\slugcomment{Submitted to the The Astrophysical Journal}
\slugcomment{Send proofs to:  J. Melendez}

\begin{abstract}

We have determined magnesium isotopic ratios ($^{25,26}$Mg/Mg) in metal-poor 
($-2.6 \leq$ [Fe/H] $\leq -1.3$)
halo dwarfs employing high S/N (90-280) high spectral resolution
(R = 10$^5$) Keck HIRES spectra.
Unlike previous claims of an important contribution from intermediate-mass
AGB stars at low metallicities, we find that the rise of the AGB contribution 
in the Galactic halo did not occur until intermediate metallicities ([Fe/H] $\gtrsim -$1.5).
\end{abstract}


\section{Introduction}

Magnesium is composed of three stable isotopes $^{24}$Mg, $^{25}$Mg and $^{26}$Mg,
which can be formed in massive stars (e.g. Woosley \& Weaver 1995; hereafter WW1995).
The lightest isotope is formed as a primary isotope from H, 
while $^{25,26}$Mg are formed as secondary isotopes.
The heaviest Mg isotopes are also produced in intermediate-mass 
AGB stars (Karakas \& Lattanzio 2003), 
so the isotopic ratios $^{25,26}$Mg/$^{24}$Mg increase
with the onset of AGB stars.
Therefore, Mg isotopic ratios in halo stars 
could be used to constrain the rise of AGB stars in our Galaxy. 

It is important to know when AGB stars begin to enrich the halo
in order to disentangle the contribution of elements produced by
intermediate-mass stars from those produced by masssive stars. 
For example, the high nitrogen abundances observed in metal-poor stars 
can be explained by fast-rotating massive stars 
(Chiappini, Matteucci \& Ballero 2005; Chiappini et al. 2006) 
or alternatively by intermediate-mass stars, although 
the latter option may be unlikely because those stars 
may not have had time to enrich the halo due to their longer lifetime.

Mg isotopic abundances can be obtained from the analysis
of MgH lines in cool stars. 
After the early work of Boesgaard (1968) and Bell \& Branch (1970),
other studies have increased the coverage in metallicity 
down to [Fe/H] = $-$1.8 (Tomkin \& Lambert 1980; Lambert \& McWilliam 1986; 
Barbuy 1985, 1987; Barbuy, Spite \& Spite 1987; McWilliam \& Lambert 1988; 
Gay \& Lambert 2000). 

In order to reach lower metallicities ([Fe/H] $< -$2), very metal-poor
cool dwarfs have to be discovered. This work was undertaken
by Yong \& Lambert (2003a,b), who found a number of metal-poor ([Fe/H] $< -$2)
cool dwarfs (\teff $<$ 5000 K) useful for Mg isotopic studies.
Employing that sample, Yong, Lambert \& Ivans (2003, hereafter YLI2003) 
were able to study $^{25,26}$Mg/$^{24}$Mg ratios down to [Fe/H] = $-$2.5.
Surprisingly, they found metal-poor stars with 
relatively high $^{25,26}$Mg/$^{24}$Mg ratios, suggesting thus
an important contribution by intermediate-mass AGB stars even
at such low metallicities (Fenner et al. 2003, hereafter F2003).

In this work, we determine Mg isotopic ratios in
cool halo dwarfs and constrain the rise of intermediate-mass AGB stars
by comparing the observed ratios with chemical evolution models.

\section{Sample stars and Observations}

The sample was selected from previous spectroscopic analyses 
of metal-poor cool dwarfs (Yong \& Lambert 2003a,b).
Five metal-poor stars were chosen
covering the range $-2.6 \leq$ [Fe/H] $\leq -$1.3:
G 69-18 (LHS 1138), G 83-46 (LHS 1718), G 103-50, 
G 63-40 (LHS 2765) and the well-known moderately metal-poor dwarf HD 103095,
with [Fe/H] = $-$2.6, $-$2.6, $-$2.2, $-$1.9 and $-$1.3, respectively.

The observations were obtained with HIRES (Vogt et al. 1994) at the 
Keck I telescope. 
The first set of spectra was taken in August 2004, just a few days after
a HIRES upgrade, taking thus advantage of improvements in efficiency,
spectral coverage and spectral resolution. A resolving power of 
R $\approx 10^5$ was achieved using a 0.4\arcsec-wide slit.
Additional observations were obtained in November 2004 and June 2005.

The spectral orders were extracted with MAKEE\footnote{MAKEE was developed
by T.A. Barlow specifically for reduction of Keck HIRES data.  It is
freely available at http://www2.keck.hawaii.edu/inst/hires/data\_reduction.html}
and IRAF was used for further data reductions (Doppler correction,
continuum normalization and combining spectra).

Two sample stars (G 83-46 and G 103-50) turned out to be
double lined stars, with G 103-50 being a spectroscopic binary
(Latham et al. 1988).
These two stars were discarded from the analysis.

\section{Atomic and Molecular Data}

Three wavelength regions at 5134.6, 5138.7 and 5140.2 \AA\ are
usually employed to determine the isotopic abundance ratios
$^{25,26}$MgH/$^{24}$MgH (e.g. McWilliam \& Lambert 1988; 
YLI2003). For these regions we adopted laboratory FTS measurements
of the isotopic $^{24,25,26}$MgH lines obtained by
Bernath, Black \& Brault (1985).

In addition, outside the recommended regions, laboratory wavenumbers
for $^{24}$MgH were taken from Bernath et al. (1985), and the
corresponding $^{25,26}$MgH line positions were computed by
adding the theoretical isotopic shifts to the laboratory $^{24}$MgH wavenumbers.
The $^{25,26}$MgH isotopic shifts were calculated using the
relative reduced mass of $^{25,26}$MgH to $^{24}$MgH.

The energy levels were calculated using molecular constants by 
Shayesteh et al. (2004) and we adopted a dissociation energy of 
1.27 eV (Balfour \& Lindgren 1978). Oscillator strengths were obtained 
from transition probabilities given by Weck et al. (2003).

Molecular C$_2$ lines are also present in the same region, so a line list
of C$_2$ lines was also implemented. Laboratory wavenumbers were
taken from  Amiot (1983) and Prasad \& Bernath (1994).
The rotational strengths (H\"onl-London factors) were computed
following Kovacs (1961), and we adopted an oscillator band strength of 
$f_{00}$ = 0.03 (see Grevesse et al. 1991).
Excitation potentials were computed using the molecular
constants by Prasad \& Bernath (1994) and a dissociation energy of 
6.297 eV (Urdahl et al. 1991) was adopted.

Atomic lines present in the region were also included. 
The initial line list was based in the work of Barbuy (1985), 
and lines were added or discarded based on spectral synthesis 
of both the Sun and Arcturus spectra.

In previous works the macroturbulence has been determined mainly 
using two lines: \ion{Ni}{1} 5115.4 \AA\ and \ion{Ti}{1} 5145.5 \AA\ 
(e.g. McWilliam \& Lambert 1988; YLI2003). For the two more metal-poor
stars in our sample these lines became too weak, so we
additionally used lines of \ion{Fe}{1} and \ion{Ca}{1} present
in the 5569-5601 \AA\ region.

For the atomic lines we adopted transition probabilities from
the NIST database\footnote{http://physics.nist.gov/PhysRefData/ASD/};
astrophysical {\it gf}-values were derived when no entry was available.

\section{Spectral Synthesis Analysis}

The stellar parameters (\tsin, log g, [Fe/H], v$_t$) 
were initially adopted from Mel\'endez \& Barbuy (2002) for HD 103095,
and from Yong \& Lambert (2003b) for the other two stars. A check of the
stellar parameters was done employing the IRFM \teff calibrations by
Ram\'{\i}rez \& Mel\'endez (2005), Hipparcos parallaxes, Y$^2$ isochrones
(Demarque et al. 2004) and our HIRES spectra. E(B-V) was estimated
both using interstellar \ion{Na}{1} D lines and reddening maps ($\S$4.1 of Mel\'endez et al. 2006).
Reasonable agreement was found with respect to the stellar parameters
given in the above references. Our final adopted values are given in Table 1.

Once the stellar parameters were set, the macroturbulence was determined
employing the \ion{Ni}{1} 5115.4 \AA\ and \ion{Ti}{1} 5145.5 \AA\ lines,
as well as \ion{Fe}{1} and \ion{Ca}{1} lines around 5569-5601 \AA.

The contribution of C$_2$ lines was constrained by spectral synthesis of the
weak feature around 5135.7 \AA, which is a blend of C$_2$ lines
(5135.57 and 5135.69 \AA) sometimes blended with an unidentified line 
in the red side (Gay \& Lambert 2000).
Fortunately the observations are of such high resolution that it is
possible to constrain the contribution of C$_2$ employing the
blue side of this feature, imposing thus an upper limit to blends by C$_2$ lines.

The Mg isotopic ratios were determined using spectral synthesis.
After the first trials it was clear that the $^{25,26}$Mg isotopic 
ratios were lower than 5\%, i.e., much lower than the 
terrestrial ratios (79:10:11). The computed synthetic spectra
have isotopic ratios ranging from $^{24}$Mg:$^{25}$Mg:$^{26}$Mg=100:0:0 to 90:5:5. 

Initially the Mg isotopic ratios were determined by an eye-fit of the
synthetic spectra to the HIRES observed data of the three recommended regions
(see $\S$3). The results are shown in Table 1.
After the fits by eye were completed we performed a $\chi^2$ fit by computing
$\chi^2$ = $\Sigma$($O_i - S_i$)/$\sigma^2$, where $O_i$ and $S_i$ represents
the observed and synthetic spectrum, respectively, and $\sigma$ = (S/N)$^{-1}$.
As an example of the $\chi^2$ fits we show in Fig. 1 the fits for
the recommended region at 5140.2 \AA\ in the most metal-rich 
and most metal-poor stars of our sample.
The results of the $\chi^2$ fits are shown in Table 1. As can be seen,
the eye fit compares well to the $\chi^2$ fit. The errors given in Table 1
are due to statistical errors (the standard deviation between the isotopic ratios
of the three recommended regions) and systematic errors of 1\% (due to errors
in the atmospheric parameters, see e.g. YLI2003).
Our results for HD 103095 ($^{24}$Mg:$^{25}$Mg:$^{26}$Mg = 94.8:2.4:2.8) 
compare very well with previous visual (eye-fit) 
determinations in the literature.
For HD 103095 both Tomkin \& Lambert (1980) and Barbuy (1985) 
obtained isotopic ratios of 94:3:3.
More recently Gay \& Lambert (2000) determined 93:4:3.

\section{Discussion}

Here we discuss how our isotopic ratios compare with chemical evolution models.
We compare only the ratio $^{26}$Mg/$^{24}$Mg, since the isotopic ratio
for $^{25}$Mg is more uncertain due to the smaller isotopic shift.

In Fig. 2 (left panel) we compare our results with the models computed
by F2003, both including and neglecting the contribution of
intermediate-mass AGB stars. Another model (Alib\'es, Labay \& Canal 2001, hereafter ALC2001)
which includes only massive stars is also shown.
A comparison with other models (Ashenfelter, Mathews \& Olive 2004, hereafter AMO2004;
Goswami \& Prantzos 2000) is shown in the right panel.
As can be seen, our low isotopic ratios can be explained mostly
by massive stars, thus we find no need to invoke the contribution of 
intermediate-mass AGB stars at low metallicities. HD 103095 lies slightly
above the predicted F2003 curve of massive stars nucleosynthesis (although
it is in perfect agreement with the ALC2001 model), so this may
indicate that at [Fe/H] $\gtrsim -$1.5 the contribution from AGB stars begins.

The high isotopic ratios found in metal-poor stars by YLI2003
were interpreted by F2003 as an important contribution of
intermediate-mass AGB stars at low metallicities. However, most of the stars with high 
isotopic $^{25, 26}$Mg/Mg ratios in YLI2003 are not
bona fide halo dwarfs. We computed the probability of halo membership following
Bensby, Feltzing \& Lundstr\"om (2003) 
and found that a fraction of the metal-poor stars in
YLI2003 are actually thick disk stars. Furthermore,
some halo stars have abundance anomalies (e.g. CH stars) or their
spectra are abnormal (e.g. double lined), and they should be removed 
for a fair comparison with chemical evolution models. 

After eliminating the probable thick disk stars, as well as 
halo stars with anomalies, we find that only 4 bona fide halo dwarfs remain
from the YLI2003 sample:
G 39-36, LHS 3780, G 113-40 and G 86-39. 
As can be seen in Fig. 2 (right panel), the results of YLI2003
are in excellent agreement with ours\footnote{YLI2003 report that
G 39-36 was observed with a resolving power of R = 60 000, but their
other 3 metal-poor dwarfs were observed with a lower R = 35 000.
For G 39-36 we adopted the typical error of 3\% quoted by YLI2003,
but for the other 3 dwarfs in their sample we increased the error to 4\% 
due to the lower resolving power of the observations}.
We have done a similar exercise with the sample of Gay \& Lambert (2000)
and found that the only good unevolved halo star is HD 103095, which is
already included in our sample. Lambert \& McWilliam (1986) have analyzed
the metal-poor ([Fe/H] = $-$1.5) subgiant $\nu$ Ind, for which they
obtained only upper limits of $^{25, 26}$Mg/Mg $\leq$ 3\%.

Thus both our results and YLI2003 suggest a
small (or none) $^{25, 26}$Mg contribution of intermediate-mass AGB stars to the Galactic halo. 
Perhaps the $^{25, 26}$Mg yields from AGB stars are lower
than in current models (Karakas \& Lattanzio 2003).
If this is the case, then intermediate-mass 
AGB stars can not be invoked to explain the possible variation of the 
fine-structure constant $\alpha$ (AMO2004). 
The chemical evolution models used to 
explain variations in $\alpha$ require an ad-hoc AGB-enhanced IMF 
in order to produce large amounts of $^{25, 26}$Mg (AMO2004).
If the correct yields are lower than present calculations,
then much larger ad-hoc modifications to the IMF would 
be required.

Calculations of Karakas \&  Lattanzio (2003) show that 
the AGB stars that contribute significant amounts of $^{25, 26}$Mg are stars with
initial masses of 3-6 M$_\odot$. Since these stars have lifetimes
considerably shorter than the age of the universe, they can be
used to constrain the timescale for the formation of the
Galactic halo. According to the 
Padova evolutionary tracks\footnote{http://pleiadi.pd.astro.it/}
the lifetime of 3-6 M$_\odot$ metal-poor stars ([Fe/H] = $-$1.5)
are 0.1-0.3 Gyr, so the halo timescale formation should be
of the order of 0.3 Gyr. This short timescale probably explains why
recent studies of age spread in Galactic globular clusters have 
shown that most clusters from intermediate to low metallicity are 
coeval within the uncertainties (e.g. Rosenberg et al. 1999; De Angeli et al. 2005).

\section{Conclusions}

We have shown that the $^{25, 26}$Mg/Mg ratios in halo
dwarfs are low and that there is no need to invoke a
contribution from intermediate-mass AGB stars at low metallicities. 

Further high S/N high spectral resolution observation
of a larger sample will help constrain the rise of AGB stars 
in the Galaxy and will be useful to better constrain the
formation timescale of the Galactic halo.

\acknowledgements

JM thanks C. Chiappini for useful discussions and 
D. Yong for providing data to check our MgH line list,
as well as for useful discussions on Mg isotopic ratios. 
JGC is grateful for partial support to NSF grant  AST-0507219.
We  have made use of data from the SIMBAD database operated at CDS.
The authors wish to recognize and acknowledge the very significant 
cultural role and reverence that the summit of Mauna Kea has always had 
within the indigenous Hawaiian community. 
We are most fortunate to have the opportunity to conduct observations from this mountain.

\clearpage

\begin{deluxetable}{lllllllllllll}
\footnotesize
\tablecaption{Atmospheric Parameters and Mg isotopic ratios}
\tablehead{
\colhead{ID} & \colhead{$E_{B-V}$} & \colhead{\teff} &
\colhead{log $g$} & \colhead{[Fe/H]} & \colhead{v$_{mic}$} & \colhead{v$_{mac}$} &
 \multicolumn{2}{c}{$\chi^2$ fit*} & \multicolumn{2}{c}{eye fit*} \\
\colhead{}  & \colhead{(mag)}  & \colhead{(K)} &
\colhead{(dex)}  & \colhead{(dex)} & \colhead{(km s$^{-1}$)} & \colhead{(km s$^{-1}$)} &
\colhead{$^{25}$Mg (\%)} & \colhead{$^{26}$Mg (\%)} &
\colhead{$^{25}$Mg (\%)} & \colhead{$^{26}$Mg (\%)} }
\startdata
HD 103095 & 0.000 & 5010 & 4.60 &-1.35 & 0.5 & 3.0 & 2.4$\pm$1.3 & 2.8$\pm$1.6 & 3.5$\pm$1.5 & 3.3$\pm$1.5 \\
G 63-40   & 0.005 & 4686 & 4.81 &-1.86 & 0.3 & 2.0 & 0.5$\pm$2.2 & 1.2$\pm$1.4 & \nodata     & 1.0$\pm$2.0\\
G 69-18   & 0.030 & 4480 & 4.75 &-2.60 & 0.3 & 1.5 & 0.9$\pm$2.0 & 1.5$\pm$2.2 & \nodata     & 1.0$\pm$2.0\\
\enddata
\tablenotetext{*}{Mg isotopic ratios are given with respect to $^{24}$Mg+$^{25}$Mg+$^{26}$Mg
and are expressed as percentages.
}
\label{tabiso}
\end{deluxetable}

\begin{figure}
\epsscale{}
\plottwo{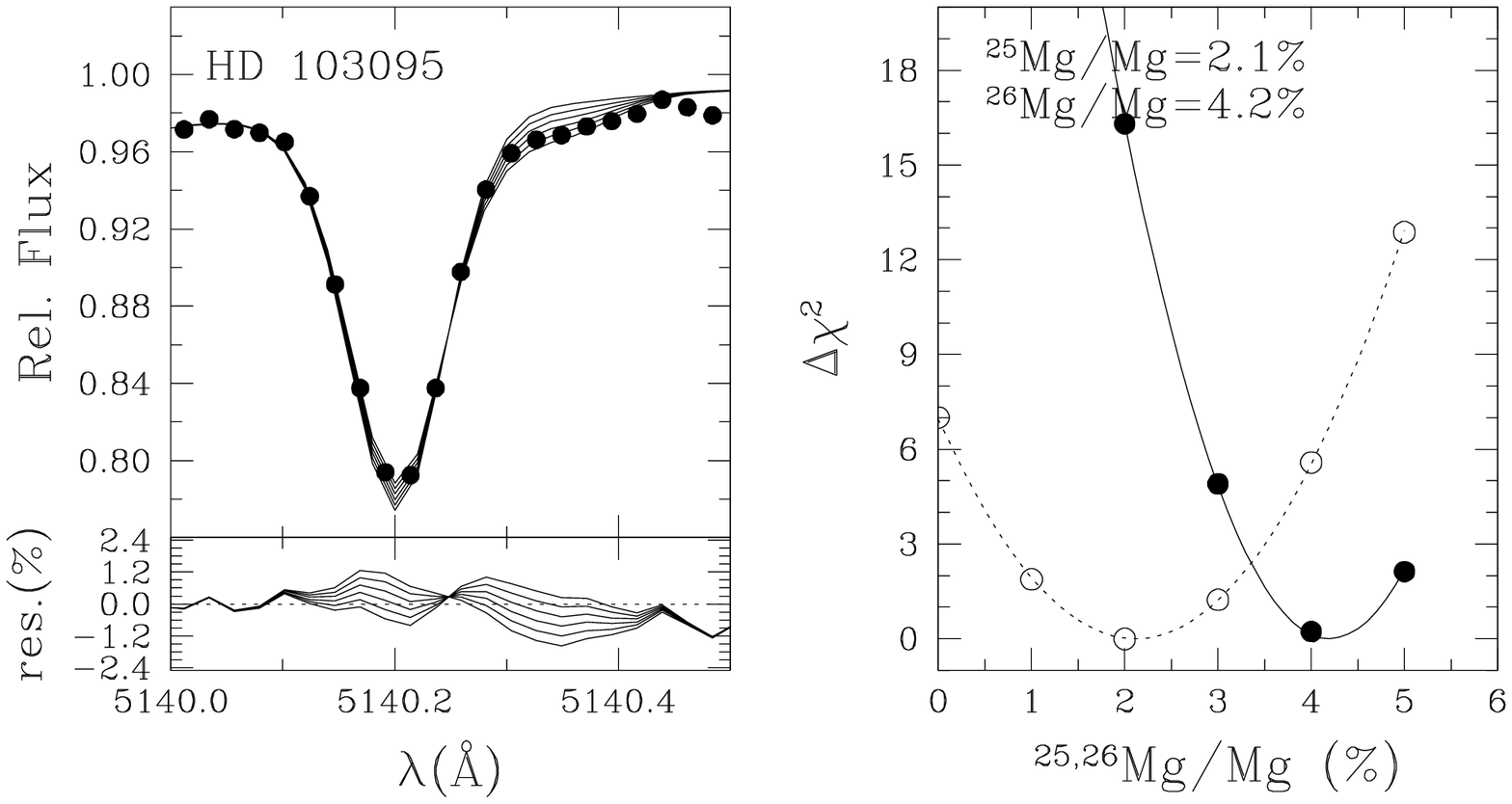}{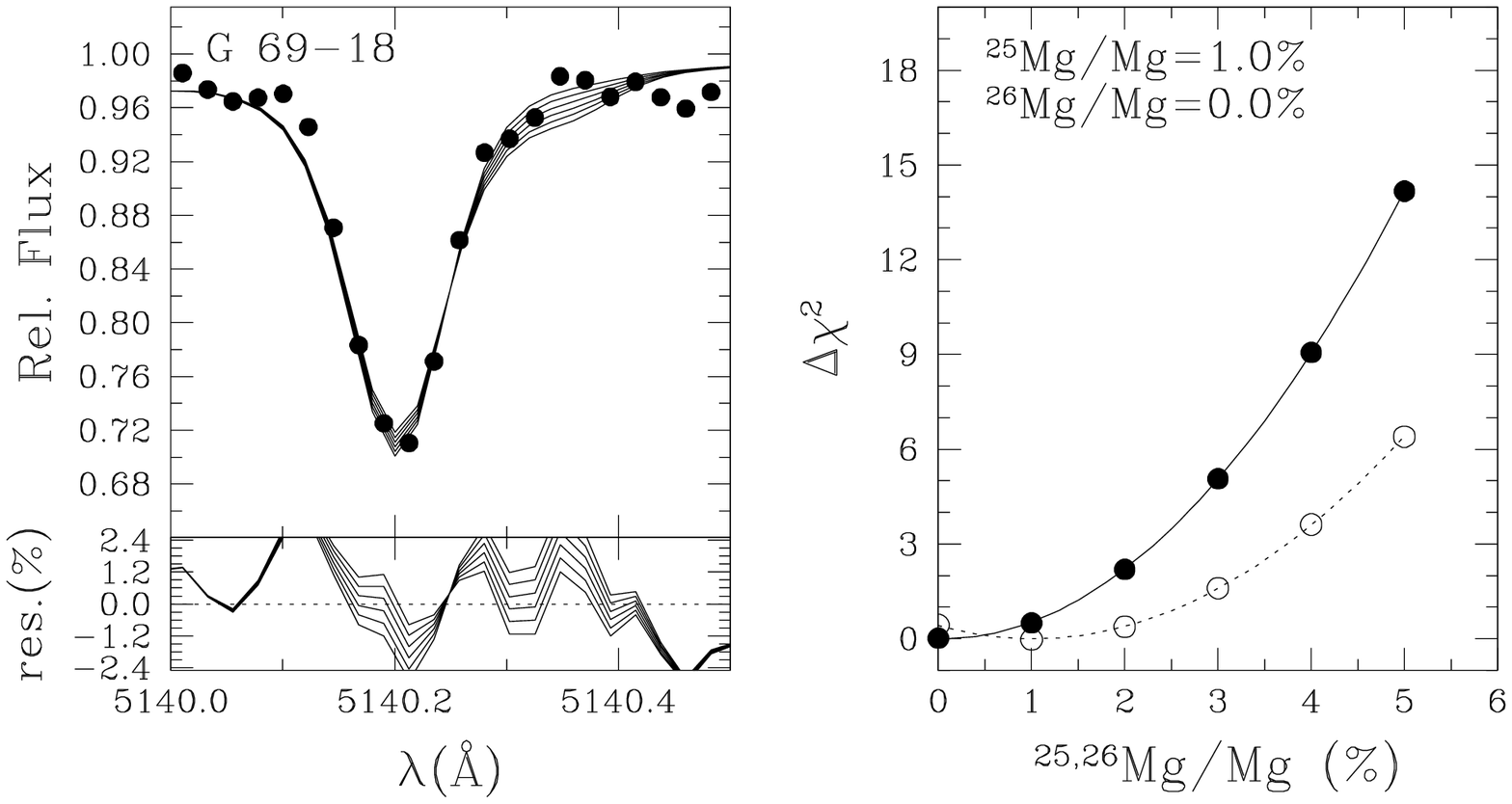}
\caption{Fits for the 5140.2 \AA\ region in the stars HD 103095 and G 69-18.
Observed spectra are represented with filled circles, and synthetic
spectra with solid lines. The calculations were performed for
$^{25,26}$Mg/Mg ratios of 0-5\%. The relative variation of the $\chi^2$ fits
are shown as a function of the isotopic abundance.
}
\label{mghfit}
\end{figure}

\begin{figure}
\epsscale{}
\plottwo{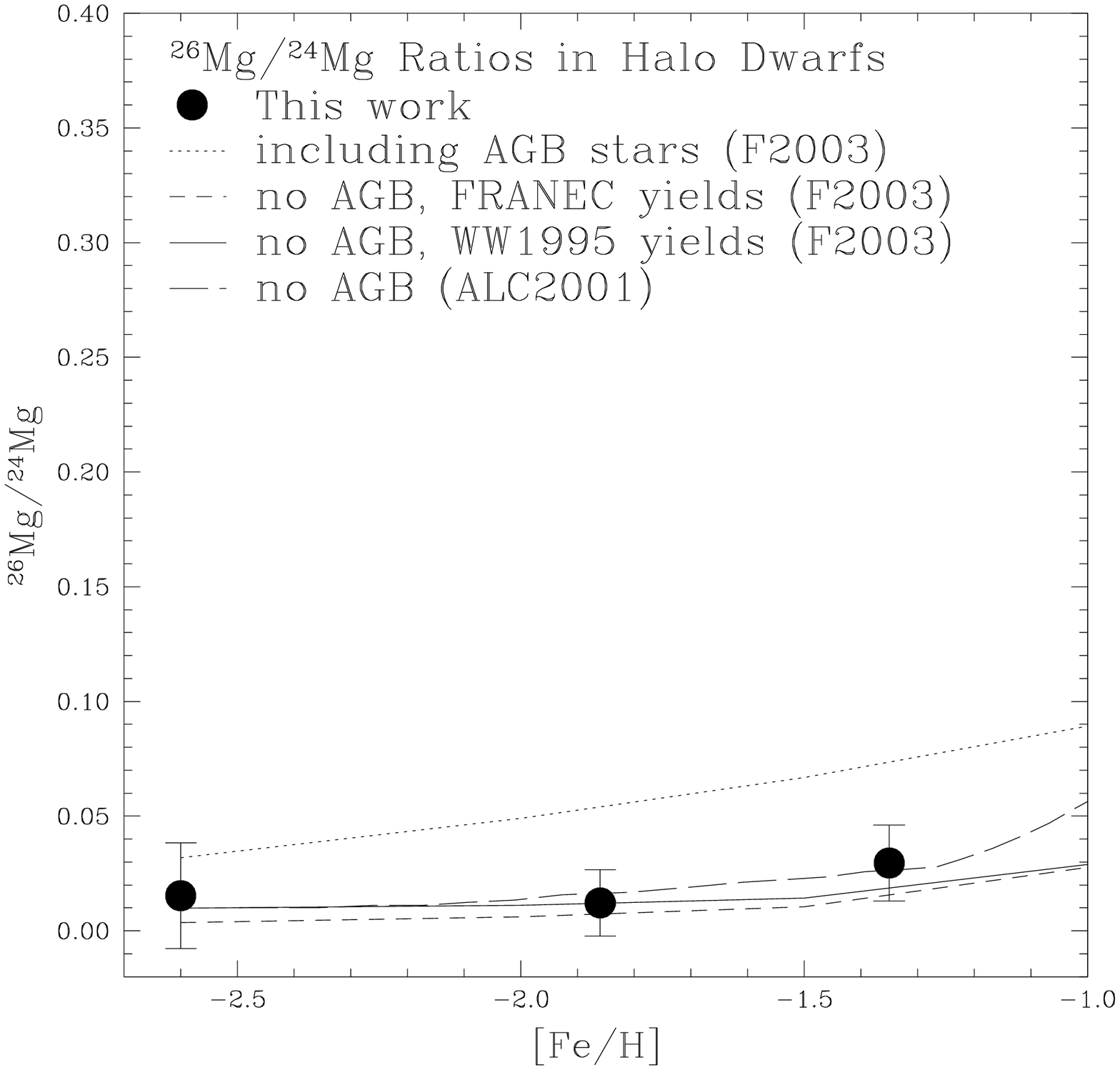}{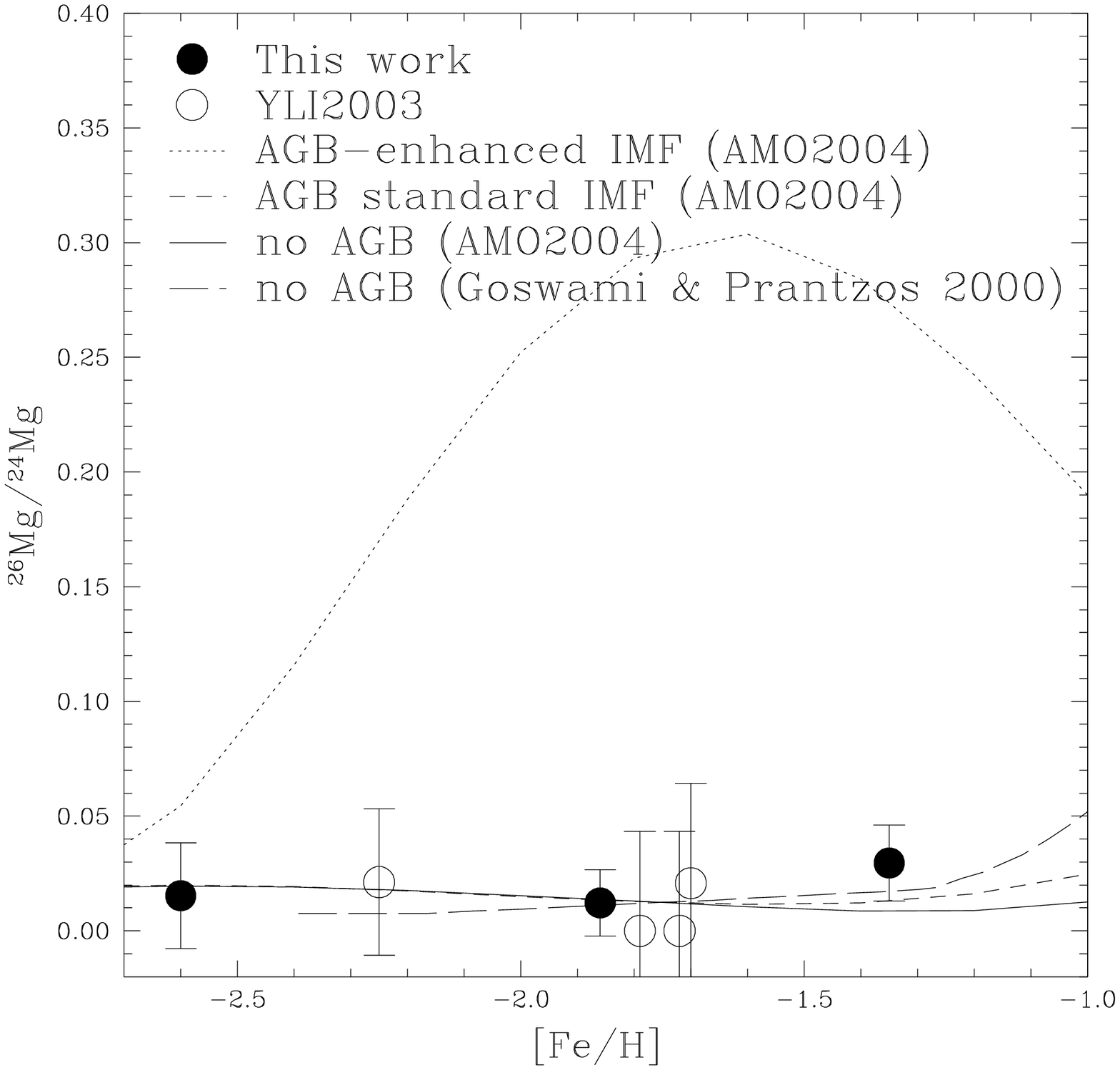}
\caption{$^{26}$MgH/$^{24}$MgH as a function of [Fe/H] in halo dwarfs.
Filled and open circles represent our results and those by
YLI2003, respectively. All models include yields of massive stars
(mostly by WW1995). Models including massive stars and intermediate-mass
AGB stars (F2003 and AMO2004) are also shown. 
Note that AMO2004 extrapolated AGB yields for Z $<$ 0.004, 
so their results may be unrealiable.
The model that agrees better
with the observed data is the ALC2001 model, which does not include 
intermediate-mass stars.
}
\label{mghfeh}
\end{figure}

\end{document}